\documentclass{aemj}
\usepackage{cite}
\usepackage{amsmath,amssymb,amsfonts}
\usepackage{algorithmic}
\usepackage{graphicx}
\usepackage[export]{adjustbox}
\usepackage{textcomp}
\def\BibTeX{{\rm B\kern-.05em{\sc i\kern-.025em b}\kern-.08em
    T\kern-.1667em\lower.7ex\hbox{E}\kern-.125emX}}
\begin{document}

\title{Radio Interferometer with Simple antennas}
\author{Akash Kulkarni\authorrefmark{1}}
\address[1]{School of Electronics and Communication Engg., KLE Technological University,BVB Campus, Vidyanagar, Hubli, Karnataka, India- 580031}

\markboth
{Akash Kulkarni}
{Akash Kulkarni}

\corresp{Corresponding author: Akash Kulkarni (e-mail: akash.kulkarni@kletech.ac.in)}

\begin{abstract}
A Radio interferometer comprises several antennas, spared over a large area. Say ALMA(Atacama Large Millimeter/submillimeter Array), VLA(very large array), VLBA(Very Long Baseline Array), GMRT(Giant Metrewave Radio Telescope), MWA( Murchison Widefield Array), EHT(Event Horizon Telescope), and the SKA(Square Kilometer Array), the name itself speaks about square-kilometres of area. Most radio observatories are constructed or constitute giant dish antennas, and few constitute an extensive array of antennas. However, what if a simple antenna like Dipole, Loop or Yagi-UDA is considered an element of an interferometer? Then how does it affect the visibility of the instrument? Yes, it will be less, but how weak?
Furthermore, what is the math to reach it? These questions pushed for this work. Here, one can find the detailed derivation starting from a simple Young's double slit experiment to a radio interferometer intensity distribution in terms of the Gain of the antenna element. This literature aided in understanding the interferometer of yagi antennas of Gain 11dBi, resulting in a visibility of 0.0714. This clarity was insignificant in the current work. Hence using this work, one can design and construct a suited interferometer for their requirement.
\end{abstract}


\titlepgskip=-15pt

\maketitle

\section{Introduction}
\label{sec:introduction}
\PARstart{K}{arl} Jansky was the first person to report the existence of cosmic radio emissions in 1932. By 1937,
Grote Reber, using a 9.5-meter di antenna, could map the sky for 160 MHz and presented
variations in the radiation intensity from different sky parts. Hence started the new era of space study
of radio astronomy, and several antenna engineers and astrophysicists constructed radio telescopes
to study the sky [1]. Large dish antennas are constructed to observe and study astronomical objects
with higher resolution [12], but to avoid the high risk of mechanical instability and substantial
construction costs, interferometry, an aperture synthesis method, is used to construct the ALMA like
telescopes [6].
Low-frequency astronomy has taken wind recently due to its applications in studying red-shift, cosmic
dawn and Interstellar medium [3]. Low-frequency observatories like GMRT, LWA, VLA, WSRT, and SKA
use dish antennas [2,7,9,10,14], and MWA uses dipole array antennae as the interferometer
elements[8]. Gauribidanur solar observatory uses a log-periodic antenna array as the primary radio
element [4]. So all observatories use either dish antennas or array antennas as the interferometer
elements.

Further, this presents an opportunity to know what happens if one uses a simple antenna, say a Yai-
Uda, a spiral or any other antenna type, as these antennas are easy to design at lower frequencies.

Hence it is essential to understand interferometry in the light of antenna parameters. The radio
interferometer is a development over a young double-slit experiment. Considering the experimental
results of Pearson et al. 2018 on wide-slit effects on the ideal experiment presents the opportunity to
cross-reference the optics to antenna parameters [13].
As the available literature does not clarify the effect of using an antenna and its varying parameters,
this work aims to derive the intensity expression for an interferometer in terms of physical antenna
parameters like the gain of the antenna and spacing between two elements, i.e. the Baseline and path
difference of the signal flow. So one can construct an interferometer with any antenna as an element
and get enhanced results. This literature aims to study the summing interferometry of the coherent
source of an angular width located far away.
In section 2, we shall reconstruct intensity expression in antenna parameters. The new expression
shall be investigated for results in section 3. We shall see the conclusion in section 4. 

\section{Theory and Derivation}
\label{sec: Theory and Derivation}
\Figure[t!](topskip=0pt, botskip=0pt, midskip=0pt)[max size={\textwidth}{\textheight}]{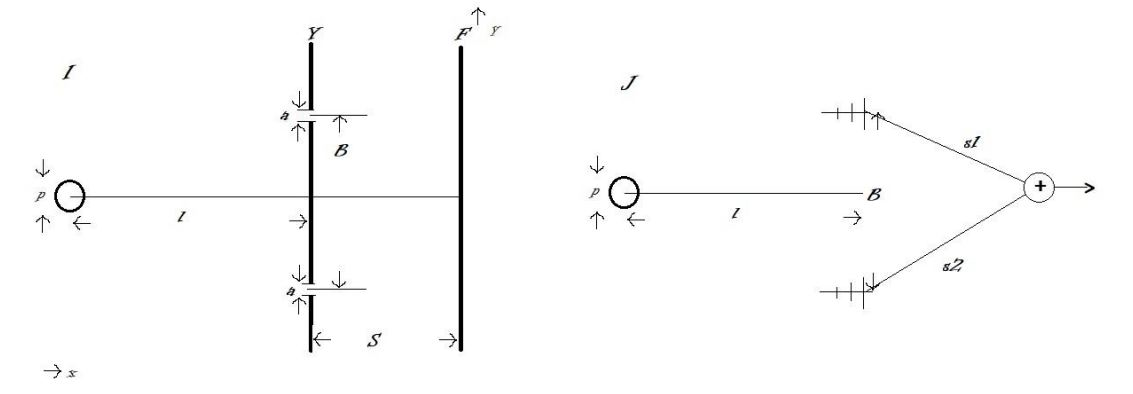}
{Correlating representation of young's double slit experiment and the radio interferometers.\label{fig1}}

Young's double slit is a base experiment for the radio interferometer. Figure 1 depicts the correlation between both. In young's experiment, a distant source's light at a certain length is $l$ and is passed through two slits separated by a distance $B$. In the ideal case, the slit width is identical; here, it is of $a$ in size. The pattern is observed on a frame at a distance $S$. In a similar sense, it is observed in radio interferometers. Let a source be at a distance $l$ and the separation between the antenna elements be $B$, also called baseline length. In figure 1, $I$ depicts a standard young double slit experiment, and $J$ is a schematic representation of a two-element radio interferometer.
\subsection{Intensity function}

The Intensity function of a young double-slit experiment is a function of slit width $a$, centre-to-centre separation of the slit $B$, and the distance to the screen $S$, with a general assumption that $S>>a, B$. To this assumption, even the source's aperture size also adds up as $S>> p^2/\lambda$, where $p$ is the aperture size, and $\lambda$ is the wavelength of the light. Hence the expression is presented along the y-axis as given by 
\begin{equation}
    I(y) = {\rm sinc}^2\left( \frac{kay}{2S}\right)\cos^2\left(\frac{kBy}{2S}\right)
\end{equation}
This expression is extended to consider the aperture size of the source given by,
\begin{equation}
    I(y) = \frac{1}{2}{\rm sinc}^2 \left(\frac{kay}{2S}\right)\left[1+{\rm sinc}\left(\frac{k\Phi B}{2}\right)\cos\left(\frac{kBy}{S}\right)\right]
\end{equation}
And $\Phi = p/l$ where $l$ is the distance between the source and the slit frame $y$ as depicted in figure 1.

Over comparison, From figure 1, a few variables are the exact representation between $I$ and $J$. $B$ is the centre-to-centre separation between the slits or the Baseline length. Angular aperture $\Phi$ of the source and wavenumber $k$ are the same in both experiments. At the same time, $S$ the mean distance between the light source and screen and $a$ the width of the slit, are not precisely replaced$!$ Now let these variables be explored. 
\subsection{Mean Distance }
The mean distance is the spacing between the slit apparatus and the screen on which the pattern is observed. So $S$ is the average length the light rays take to reach the screen.

Consider $s1$ and $s2$ be the electrical signal path length of the signal collected by the antennas. In correlation, if we consider letting the $s1$ and $s2$ also be the path distance of light rays from the slits. Then using the young double-slit calculations we have,
\begin{equation}
    s1 = \sqrt{S^2+\left(y-\frac{B}{2}\right)^2}
\end{equation}
\begin{equation}
    s2 = \sqrt{S^2+\left(y+\frac{B}{2}\right)^2}
\end{equation}
i.e.
\begin{equation}
    s2^2 - s1^2 = 2yB
\end{equation}
\begin{equation}
    s2-s1 = \frac{2yB}{s2+s1}
\end{equation}
As $S$ is the average of both the path lengths of the light rays. We have,
\begin{equation}
    s2+s1 =2S
\end{equation}
And taking the difference of the length as $\Delta s$. We have 
\begin{equation}
    \Delta s = \frac{yB}{S}
\end{equation}

The electric path along the different cables tends to produce a phase difference at the summing end of the system. This similarity can be brought to the expression 8 in similarity to the difference in path length.

i.e.
\begin{equation}
    \delta =\frac{nkyB}{S}
\end{equation}

i.e.
\begin{equation}
    S =\frac{nkyB}{\delta}
\end{equation}
Where $k$ stands for wavenumber. $n$ is a positive integer to be chosen so that $S>>B$, the initial assumption of the young double-slit experiment[5].  
\subsection{Slit width}
The slit width is an essential part of young's double slit experiment, as we have an interference pattern only because of these two slits. The width of the slit $a$ or aperture of the slit, the one which controls the flow of light. The larger the aperture, the more light passes through the slit $i.e.$ more flux pass-through. Here it is comparable to the effective aperture of the antenna.
The effective aperture of an antenna is the amount of collecting area for the radiation. $i.e.$, as the effective aperture increases,  the amount of flux collected by the antenna  also increases. Hence, we shall have the area of slits as a direct replacement for an effective aperture, considering a circular slit of diameter $a$. 

$i.e.$
\begin{equation}
    A_e = \pi a^2
\end{equation}
The effective aperture is proportional to the antenna's gain and the wavelength's square.
\begin{equation}
    A_e = \frac{\lambda ^2}{4\pi} G
\end{equation}
Hence that is,
\begin{equation}
    \frac{\lambda ^2}{4\pi} G = \pi a^2
\end{equation}
\begin{equation}
    a^2 = \frac{\lambda ^2}{4\pi ^2} G
\end{equation}
\begin{equation}
    a= \frac{\lambda}{4\pi} \sqrt{G}
\end{equation}

$i.e.$
\begin{equation}
    a = \frac{\sqrt{G}}{k}
\end{equation}
By substituting all these into the primary expression, we have 
\begin{equation}
    I(y) = \frac{1}{2}{\rm sinc}^2 \left(\frac{\sqrt{G}y}{2S}\right)\left[1+{\rm sinc}\left(\frac{k\Phi B}{2}\right)\cos\left(\frac{kBy}{S}\right)\right]
\end{equation}
Hence, the traditional Intensity function changes to adapt to the radio interferometer built by any arbitrary antenna with gain $G$. Further analysis and results are presented in the next section.
Antennas are of several forms, and their gain also varies from dipole antenna of 1.64dBi to more than 20dBi for aperture antenna. An array antenna has a directivity between 2 to 7.28 times the ratio of the distance between the end elements and wavelength. A yagi-uda antenna, like a broadband antenna, can present a gain between 7 to 20 dBd. And the maximum gain of a reflector antenna is given by,
\begin{equation}
    G_{max} = \frac{4\pi}{\lambda ^2} A_p
\end{equation}
For an aperture of the diameter of 100m or 500m, we have the gain of 45dBi to 58dBi at 160MHz, respectively. Hence the results are analysed and presented for the gain of 1 to 100dBi for visibility.

And last but not least, the relation between the gain and directivity of the antenna
$i.e.$
\begin{equation}
    G = \eta D
\end{equation}
The proportionality constant $\eta$ is between $0$ to $1$, and ideally, it is $1$.
\section{Results and discussion}
In this section, initially, we shall see the results of the ideal young's double experiment. The same experiment with double slits with a width and the modified expression for the intensity be tested. The visibility of the fringe is the primary concern of this investigation. In this experiment, all length measurements are measured in meters. 

The Ideal young double-slit experiment considers a narrow source slit and similar narrow double slits for interference. So considering a slit width of 0.1m and separation of 10 m at a screen distance of 100m, the visibility is 0.9558. For slit widths of 0.15m, 0.5m and 1m, visibility is 0.957,0.9734 and 1, respectively, as presented in figures 2 to 5, and for slit widths of 0.15m and separation of 15m and 20m, it is 0.9016, and 08281 is the visibility as seen in figure 6 and 7.
The same investigation is extended to a radio interferometer with an antenna gain in dBi and an operating frequency of 160MHz. Hence the visibility is plotted for various parameters like Baseline $B$, Gain $G$ and angular width of the source $\Phi$.
\begin{table}[h!]
\caption{Table of values of visibility of the firing pattern for Baseline length = 100m and various values of Gain and source size}
\label{table:1}
\begin{tabular}{|c | c | c | c | c |} 
 \hline
 $G(dBi) / \Phi $ & $ 1'$ & $ 5'$ & $ 10'$ & $ 1^\circ$ \\ 
 \hline
1 & 6.9e-2 &  2.81e-3 & 2.79e-3 & 2.01e-3\\
 \hline
11& 7.14e-2 & 4.51e-3 & 4.51 e-3&4.44e-3 \\
 \hline
21& 9.36e-2 & 4.343e-2& 4.343e-2& 4.348e-2 \\
 \hline
 31 &5.237e-1 & 4.753e-1 & 4.754e-1 & 4.757e-1 \\
 \hline
 41&1&1&1&1\\
 \hline
 51&1&1&1&1\\
 \hline
\end{tabular}
\end{table}

\begin{table}[h!]
\caption{Table of values of visibility of the firing pattern for Baseline length = 1000m and various values of Gain and source size}
\label{table:2}
\begin{tabular}{|c | c | c | c | c |} 
 \hline
 $G(dBi) / \Phi $ & $ 1'$ & $ 5'$ & $ 10'$ & $ 1^\circ$ \\ 
 \hline
1 & 2.68e-3&2.13e-3&8.4e-4 & 2.17e-4\\
 \hline
11& 6.88e-3 &6.33e-3 &5.04e-3 &4.42e-3 \\
 \hline
21& 4.528e-2& 4.474e-2 & 4.345e-2 & 4.287e-2 \\
 \hline
 31 & 4.447e-1& 4.443e-1& 4.719e-1 & 4.769e-1\\
 \hline
 41&1&1&1&1\\
 \hline
 51&1&1&1&1\\
 \hline
\end{tabular}
\end{table}
\Figure[t!](topskip=0pt, botskip=0pt, midskip=0pt)[max size={\textwidth}{\textheight}]{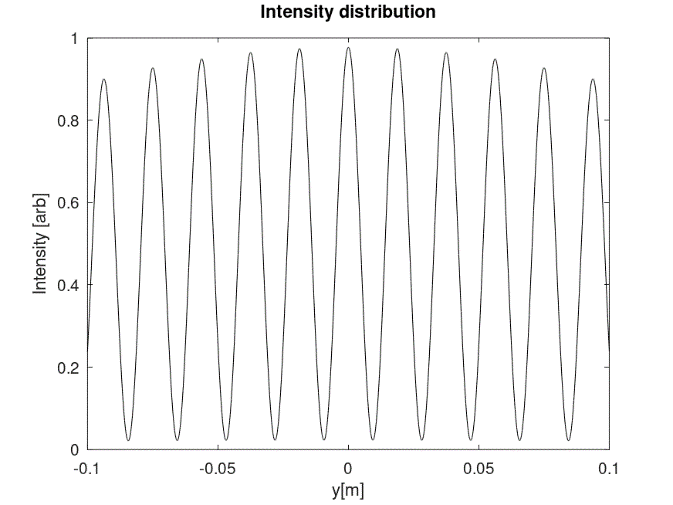}
{ Intensity pattern for $a = 0.1m B = 10m S=100m$.\label{fig2}}
\Figure[t!](topskip=0pt, botskip=0pt, midskip=0pt)[max size={\textwidth}{\textheight}]{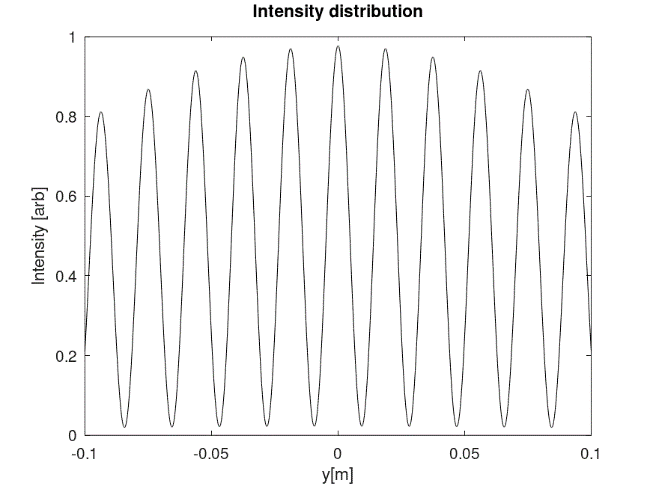}
{ Intensity pattern for $a = 0.15m B = 10m S=100m$.\label{fig3}}
\Figure[t!](topskip=0pt, botskip=0pt, midskip=0pt)[max size={\textwidth}{\textheight}]{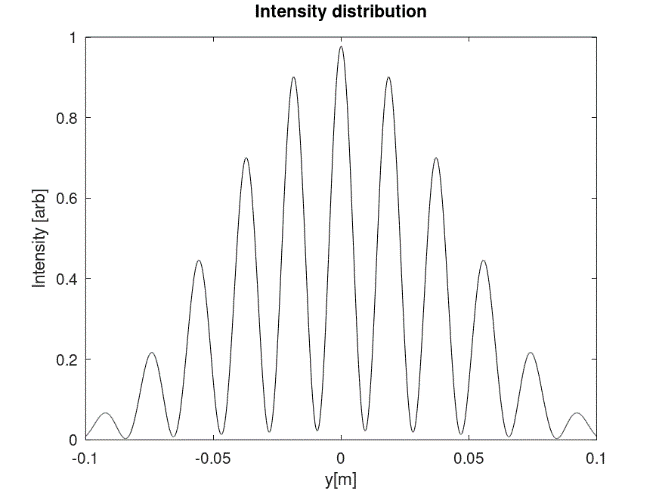}
{ Intensity pattern for $a = 0.5m B = 10m S=100m$.\label{fig4}}
\Figure[t!](topskip=0pt, botskip=0pt, midskip=0pt)[max size={\textwidth}{\textheight}]{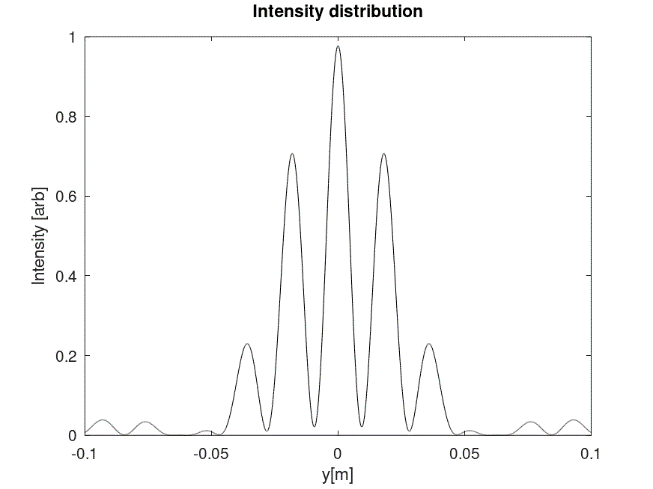}
{ Intensity pattern for $a = 1m B = 10m S=100m$.\label{fig5}}
\Figure[t!](topskip=0pt, botskip=0pt, midskip=0pt)[max size={\textwidth}{\textheight}]{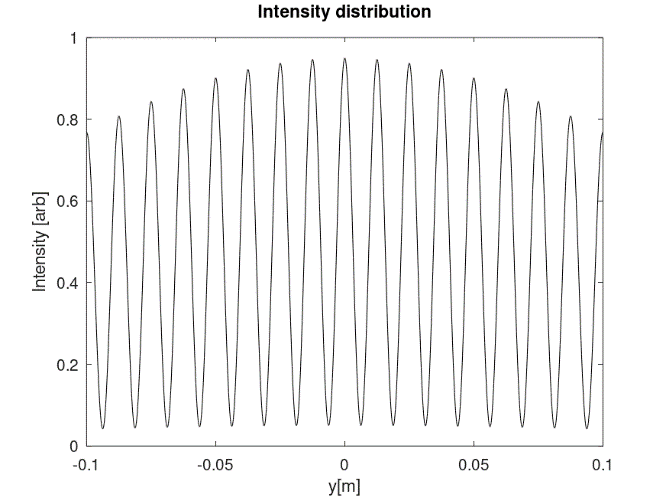}
{ Intensity pattern for $a = 0.15m B = 15m S=100m$.\label{fig6}}
\Figure[t!](topskip=0pt, botskip=0pt, midskip=0pt)[max size={\textwidth}{\textheight}]{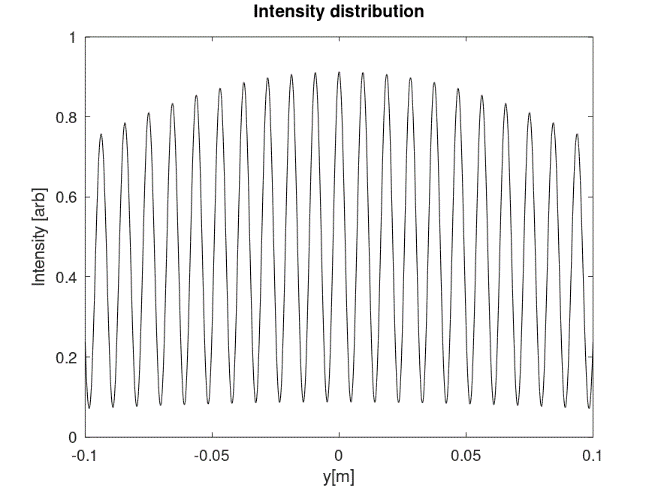}
{ Intensity pattern for $a = 0.15m B = 20m S=100m$.\label{fig7}}

Over observation, it is evident that as the gain increases, the visibility exponentially increases. As the more extensive the baseline length so the value of visibility drops. And for gain above 40dBi, as in dish or aperture antennas, the visibility reaches the full value.

\section{Conclusion}
 This experiment was used to study and understand the effect of using simple antennas instead of large dish antennas. This result concludes that, yes, it is possible, but the visibility is low, so the ability to differentiate the peaks of the interferometric pattern is challenging.  Even the antenna parameters considered here could be made more precise. By including the antenna losses, its front-end  circuitry would be crucial. These antenna specifications are particular to every antenna system design. Hence as one would design their antenna element for a particular interferometer, they can substitute those values and bring out realistic calculations for fringes and the visibility value.
\EOD
\end{document}